\documentclass[12pt]{article}
\usepackage{amssymb,amsmath,epsfig}

\begin{document}

\title{\bf Viscous Dark Energy in $f(T)$ Gravity}
\author{M. Sharif \thanks {msharif.math@pu.edu.pk} and Shamaila
Rani \thanks{shamailatoor.math@yahoo.com}\\
Department of Mathematics, University of the Punjab,\\
Quaid-e-Azam Campus, Lahore-54590, Pakistan.}

\date{}
\maketitle

\begin{abstract}
We study the bulk viscosity taking dust matter in the generalized
teleparallel gravity. We consider different dark energy models in
this scenario along with a time dependent viscous model to construct
the viscous equation of state parameter for these dark energy
models. We discuss the graphical representation of this parameter to
investigate the viscosity effects on the accelerating expansion of
the universe. It is mentioned here that the behavior of the universe
depends upon the viscous coefficients showing the transition from
decelerating to accelerating phase. It leads to the crossing of
phantom divide line and becomes phantom dominated for specific
ranges of these coefficients.
\end{abstract}

\section{Introduction}

Dark energy (DE) seems to play an important role of an agent that
drives the present acceleration of the universe with the help of
large negative pressure. An effective viscous pressure can also play
its role to develop the dynamical history of an expanding universe
\cite{57}-\cite{15}. It is found \cite{15} that viscosity effects
are viable at low redshifts, which observe negative pressure for the
cosmic expansion with suitable viscosity coefficients. In general,
the universe inherits dissipative processes \cite{23}, but perfect
fluid is an ideal fluid with zero viscosity.

Although, perfect fluid is mostly used to model the idealized
distribution of matter in the universe. This fluid in equilibrium
generates no entropy and no frictional type heat because its
dynamics is reversible and without dissipation. The dissipative
processes mostly include bulk and shear viscosities. The bulk
viscosity is related with an isotropic universe whereas the shear
viscosity works with anisotropy of the universe. The CMBR
observations indicate an isotropic universe, leading to bulk
viscosity where the shear viscosity is neglected \cite{6}. Long
before the direct observational evidence through the SN Ia data, the
indication of a viscosity dominated late epoch of accelerating
expansion of the universe was already mentioned \cite{39}.

The origin of the bulk viscosity in a physical system is due to its
deviations from the local thermodynamic equilibrium. Thus the
existence of bulk viscosity may arise the concept of accelerating
expansion of the universe due to the collection of those states
which are not in thermal equilibrium for a small fraction of time
\cite{1}. These states are the consequence of fluid expansion (or
contraction). The system does not have enough time to restore its
equilibrium position, hence an effective pressure takes part in
restoring the system to its thermal equilibrium. The measurement of
this effective pressure is the bulk viscosity which vanishes when it
restores its equilibrium \cite{22}-\cite{5+}. So, it is natural to
assume the existence of a bulk viscous coefficient in a more
realistic description of the accelerated universe today.

Physically, the bulk viscosity is considered as an internal friction
due to different cooling rates in an expanding gas. Its dissipation
reduces the effective pressure in an expanding fluid by converting
kinetic energy of the particles into heat. Thus, it is natural to
think of the bulk viscous pressure as one of the possible mechanism
that can accelerate the universe today. However, this idea needs a
viable mechanism for the origin of the bulk viscosity, although
there are many proposed best fit models.

Many models have been suggested to discuss the vague nature of DE.
During the last decade, the holographic dark energy (HDE), new
agegraphic dark energy (NADE), their entropy corrected versions and
correspondence with other DE models have received a lot of
attention. The HDE model is based on the holographic principle which
states that \emph{the number of degrees of freedom in a bounded
system should be finite and has a relationship with the area of its
boundary} \cite{50}. Moreover, in order to reconcile the validity of
an effective local quantum field, Cohen et al. \cite{12} provided a
relationship between the ultraviolet (UV) and the infrared (IR)
cutoffs on the basis of limit set by the formation of a black hole.
This is given by \cite{19,6+}
\begin{equation}\label{1}
\rho_{\Lambda}=3c^{2}M^{2}_pL^{-2},
\end{equation}
where constant $3c^2$ is used for convenience, $M_{p}^2=(8\pi
G)^{-1}$ is the reduced Planck mass and $L$ is the IR cutoff. This
model has been tested by using different ways of astronomical
observations \cite{56}-\cite{54}. Also, it has been discussed widely
in various frameworks such as in the general relativity, modified
theories of gravity and extra dimensional theories
\cite{20}-\cite{48}.

The NADE model was developed in view of the Heisenberg uncertainty
principle with general relativity. This model exhibits that DE
originates from the spacetime and matter field fluctuations in the
universe. In this model, the length measure is taken as the
conformal time instead of age of the universe and its energy density
is $\rho_{\Lambda}=\frac{3n^2}{\kappa^{2}\eta^2}$ where $\eta$ is
the conformal time. The causality problem occurs in the usual HDE
model, while it is avoided here. Many people have explored the
viability of this model through different observations
\cite{56}-\cite{54,27}.

Another proposal to discuss the accelerating universe is the
modified gravity theories \cite{cc}. The $f(T)$ gravity is the
generalization of teleparallel gravity by replacing the torsion
scalar $T$ with differentiable function $f(T)$, given by
\begin{eqnarray}\label{1*}
L_T=\frac{e}{2\kappa}T+L_m \quad \Rightarrow \quad
L_{f(T)}=\frac{e}{2\kappa}f(T)+L_{m},
\end{eqnarray}
where $\kappa$ is the coupling constant and $e=\sqrt{-g}$. This
leads to second order field equations formed by using
Weitzenb$\ddot{o}$ck connection which has no curvature but only
torsion. The equation of state (EoS) parameter, $\omega=p/\rho$, is
used to explore the cosmic expansion. Bengochea and Ferraro
\cite{aa} tested power-law $f(T)$ model for accelerated expansion of
the universe. They performed observational viability tests and
concluded that this model exhibits radiation, matter and DE
dominated phases. Incorporating exponential model along with
power-law model, Linder \cite{bb} investigated the expansion of the
universe in this theory. He observed that power-law model depends
upon its parameter while exponential model acts like cosmological
model at high redshift.

Bamba et al. \cite{4} discussed the EoS parameter for exponential,
logarithmic as well as combination of these $f(T)$ models and they
concluded that the crossing of phantom divide line is observed in
combined model only. Karami and Abdolmaleki \cite{26} constructed
this parameter for HDE, NADE and their entropy corrected models in
the framework of $f(T)$ gravity. They found that the universe lies
in phantom or quintessence phase for the first two models whereas
phantom crossing is achieved in entropy corrected models. Sharif and
Rani \cite{49} described the graphical representation of k-essence
in this modified gravity with the help of EoS parameter. Some other
authors \cite{34,52} explored the expansion of the universe with
different techniques in $f(T)$ gravity. Also, the effects of viscous
fluid in modified gravity theories \cite{7}-\cite{9} are analyzed to
display accelerating expansion.

In this paper, we construct the viscous EoS parameter for different
viable DE models in the framework of $f(T)$ gravity with
pressureless matter. For this purpose, we consider a time dependent
viscous model with its constant viscous reduction to explore the DE
era in general fluid. The graphical behavior indicates the
acceleration of the universe for suitable viscous coefficients. The
scheme of paper is as follows: Section \textbf{2} provides basic
formalism and discussion about the field equations of $f(T)$
gravity. In section \textbf{3}, the viscous EoS parameter is
constructed for different DE models. Also, we discuss the graphical
behavior of this parameter for these models. The last section
summarizes the results.

\section{The Field Equations}

The $f(T)$ theory of gravity (as the generalization of the
teleparallel gravity) is uniquely determined by the tetrad field
${h^\mu}_\alpha(x)$ \cite{35}. It is an orthonormal set of
four-vector fields defined on Lorentzian manifold. The metric and
tetrad fields can be related as
\begin{equation}\label{2}
g_{\mu\nu}=\eta_{ij}h_{\mu}^{i}h_{\nu}^{j},
\end{equation}
where $\eta_{ij}=diag(1,-1,-1,-1)$ is the Minkowski metric for the
tangent space. Here we use Greek alphabets
$(\mu,\nu,\rho,...=0,1,2,3)$ to denote spacetime components while
the Latin alphabets $(i,j,k,...=0,1,2,3)$ are used to describe
components of tangent space. The non-trivial tetrad field $h_{i}$,
yielding non-zero torsion, can be written as
\begin{equation}\label{3}
h_i={h_i}^\mu\partial_\mu,\quad h^j={h^j}_\nu dx^\nu,
\end{equation}
satisfying the following properties
\begin{equation}\label{4}
{h^i}_\mu{h_j}^\mu={\delta^i}_j,\quad
{h^i}_\mu{h_i}^\nu={\delta_\mu}^\nu.
\end{equation}
The variation of Eq.(\ref{1*}) with respect to the tetrad field
leads to the following field equations \cite{34,16}
\begin{equation}\label{5}
[e^{-1}\partial_{\mu}(eS_{i}~^{\mu\nu})
+h^{\lambda}_{i}T^{\rho}~_{\mu\lambda}S_{\rho}~^{\nu\mu}]f_{T}
+S_{i}~^{\mu\nu}\partial_{\mu}(T)
f_{TT}+\frac{1}{4}h^{\nu}_{i}f=\frac{1}{2}\kappa^{2}h^{\rho}_{i}T^{\nu}_{\rho},
\end{equation}
where $f_{T}=\frac{df}{dT},~f_{TT}=\frac{d^{2}f}{dT^{2}}$.

The torsion scalar is defined as
\begin{equation}\label{6}
T=S_{\rho}~^{\mu\nu}T^{\rho}~_{\mu\nu},
\end{equation}
where $S_{\rho}~^{\mu\nu}$ and torsion tensor $T^{\rho}~_{\mu\nu}$
are given as follows
\begin{eqnarray}\label{7}
S_{\rho}~^{\mu\nu}&=&\frac{1}{2}(K^{\mu\nu}~_{\rho}
+\delta^{\mu}_{\rho}T^{\theta\nu}~_{\theta}-\delta^{\nu}_{\rho}T^{\theta\mu}~_{\theta}),\\
\label{8}T^{\lambda}~_{\mu\nu}&=&\Gamma^{\lambda}~_{\nu\mu}-
\Gamma^{\lambda}~_{\mu\nu}=h^{\lambda}_{i}
(\partial_{\nu}h^{i}_{\mu}-\partial_{\mu}h^{i}_{\nu}),\\\label{1++}
{K^{\mu\nu}}_{\rho}&=&-\frac{1}{2}({T^{\mu\nu}}_{\rho}
-{T^{\nu\mu}}_{\rho}-{T_\rho}^{\mu\nu}),
\end{eqnarray}
which are antisymmetric. The energy-momentum tensor for perfect
fluid is
\begin{equation}\label{9}
T^\nu_\rho=(\rho+p)u^{\nu}u_{\rho}-p\delta_{\rho}^{\nu},
\end{equation}
where $u^{\nu}$ is the four-velocity in comoving coordinates, $\rho$
and $p$ denote the total energy density and pressure of fluid inside
the universe.

The flat homogenous and isotropic FRW universe is described by
\begin{equation}\label{10}
ds^{2}=dt^{2}-a^{2}(t)(dx^{2}+dy^{2}+dz^{2}),
\end{equation}
where $a(t)$ is the scale factor such that $a(t)=1/1+z$ in the form
of redshift $z$. The corresponding tetrad components are
\cite{4}-\cite{52}
\begin{eqnarray}\label{11}
h^{i}_{\mu}=diag(1,a,a,a),
\end{eqnarray}
which obviously satisfies Eq.(\ref{4}). Using Eqs.(\ref{6}) and
(\ref{10}), the torsion scalar turns out in the form of Hubble
parameter $H$ as $T=-6H^2, \quad H=\frac{\dot{a}}{a}$. The
corresponding modified Friedmann equations become
\begin{eqnarray}\label{13}
-2Tf_T+f&=&2\kappa^{2}\rho,\\\label{14}
-8T\dot{H}f_{TT}+(2T-4\dot{H})f_T-f&=&2\kappa^{2}p.
\end{eqnarray}
For the realistic model, we take viscosity term which introduces the
effective pressure in the energy-momentum tensor \cite{51}, i.e.,
$T^\nu_\rho=(\rho+p_{eff})u^{\nu}u_{\rho}-p_{eff}\delta_{\rho}^{\nu}$
defined by
\begin{equation}\label{16*}
p_{eff}=p-3H\xi(t),
\end{equation}
here $\xi$ is the time dependent bulk viscosity function. To avoid
the violation of the second law of thermodynamics, $\xi(t)>0$.

The field equations (\ref{13}) and (\ref{14}) may be rewritten as
\begin{eqnarray}\label{17}
\rho_m+\rho_T=\frac{3H^2}{\kappa^2},\quad
p_T=-\frac{1}{\kappa^2}(2\dot{H}+3H^2).
\end{eqnarray}
We assume here the pressureless (dust) matter, i.e., $p_m=0$ and the
expressions for torsion contributions $\rho_T,~p_T$ and effective
pressure become
\begin{eqnarray}\label{19}
\rho_T&=&\frac{1}{2\kappa^2}(2Tf_{T}-f-T),\\\nonumber
p_T&=&-\frac{1}{2\kappa^2}(-8T\dot{H}f_{TT}+(2T-4\dot{H})f_T\\\label{20}&-&f+4\dot{H}-T)\\\label{2++}
p_{eff}&=&p_T-3H\xi(t).
\end{eqnarray}
It is noted that if we insert $f(T)=T$ in Eq.(\ref{17}) with
non-viscous case, we arrive at the usual Friedmann equations in
general relativity. The corresponding viscous EoS parameter becomes
\begin{equation}\label{21}
\omega_{eff}=-1+\frac{2 \kappa^2 (\rho_m+3H\xi(t))}{2 \kappa^2
\rho_m+2Tf_{T}-f-T}+\frac{4\dot{H}(2Tf_{TT}+f_{T}-1)}{2 \kappa^2
\rho_m+2Tf_{T}-f-T}.
\end{equation}
The phantom and quintessence regions are mostly described with the
help of constant EoS parameter such as, $ -1<\omega<-1/3$, which
corresponds to the quintessence era whereas phantom era is referred
to $\omega<-1$ and the phantom divide line is given by $\omega=-1$.
If we consider a torsion dominated universe, then Eq.(\ref{13})
reduces to
\begin{equation}\label{a}
\frac{3H^2}{\kappa^2}=\rho_T.
\end{equation}
Inserting the above value in the energy conservation equation for
torsion, it follows that $(\omega_{eff}\rightarrow \omega_T)$
\begin{equation}\label{b}
\omega_T=-1+\frac{\kappa^2 \xi(t)}{H}-\frac{2\dot{H}}{3H^2}.
\end{equation}
The EoS parameter $\omega_T$ describes a vacuum, phantom dominated
or quintessence dominated universe for $\dot{H}=\frac{3}{2}H\kappa^2
\xi(T),~\dot{H}>\frac{3}{2}H\kappa^2 \xi(T)$ or
$\dot{H}<\frac{3}{2}H\kappa^2 \xi(T)$ respectively for viscous case.
For the non-viscous case $(\xi(t)=0)$, these conditions reduce to
$\dot{H}=0,~\dot{H}>0$ or $\dot{H}<0$.

\section{Viscous Fluid and Dark Energy Models}

Viscous models have interesting insights about the evolution of the
expanding universe. Here we consider a simple time dependent bulk
viscous model as follows \cite{40,33}
\begin{eqnarray}\label{24}
\xi(t)=\xi_0+\xi_{1}\frac{\dot{a}}{a}+\xi_{2}\frac{\ddot{a}}{a}
=\xi_0+\xi_{1}H+\xi_{2}(\dot{H}+H^2),
\end{eqnarray}
where $\xi_0,~\xi_1$ and $\xi_2$ are positive coefficients. The
cosmological evolution can be explored for different values of these
coefficients \cite{33}-\cite{21+}. This bulk viscosity model is
motivated due to the terms involved, i.e., viscosity is related to
the velocity and acceleration which give the phenomenon of scalar
expansion in fluid dynamics. The viscous model having constant
$\xi_0$ and velocity term $\dot{a}$ are discussed in \cite{40}, thus
a linear combination of these two with acceleration term $\ddot{a}$
may give more physical results.

In general, the existence of viscosity coefficients in a fluid is
due to the thermodynamic irreversibility of the motion. If the
deviation from reversibility is small, the momentum transfer between
various parts of the fluid can be taken to be linearly dependent on
the velocity derivatives. This case corresponds to the constant
viscous model. When viscosity is proportional to Hubble parameter
the momentum transfer involves second order quantities in the
deviation from reversibility leading to more physical results. The
proper choices of their coefficients may lead to the crossing of
phantom divide line.

To determine the evolution of effective EoS parameter incorporating
$f(T)$ and viscous models, we assume the Hubble parameter in the
form \cite{37,37+}
\begin{equation}\label{a}
H(t)=\frac{h}{(t_s-t)^{\gamma}}.
\end{equation}
Here $h$ and $t_s$ are positive constants, the constant $\gamma$ is
either positive or negative and $t<t_s$ is guaranteed for the
accelerated expansion of the universe due to the violation of strong
energy condition $(\rho+3p\geq0)$. For $\gamma=1$, it leads to the
scale factor $a(t)=a_0(t_s-t)^{-h}$ which ends up the universe with
future finite time Big Rip singularity. Using Eq.(\ref{a}) with
$\gamma=1$, the torsion scalar becomes $T=\frac{-6h^2}{(t_s-t)^{2}}$
with $t_s-t=(z+1)^{1/h}$. Also, taking the value of $H$, the energy
conservation equation, $\dot{\rho}_m+3H\rho_m=0$ for dust matter
yields the solution
\begin{equation}\label{y}
\rho_m=\rho_{m0}(t_s-t)^{3h}=\rho_{m0}(1+z)^{3},
\end{equation}
where $\rho_{m0}$ is an arbitrary constant.

In the following, we discuss three DE models by taking into account
of the viscosity.

\subsection{The First Model}

First, we consider the following DE model \cite{26,b}
\begin{equation}\label{29}
f(T)=\beta\sqrt{T}+(1-\alpha)T,
\end{equation}
where $\alpha$ and $\beta$ are constants. For $\beta=0$, this model
leads to the teleparallel gravity. It is interesting to note that
the model (\ref{29}) is the result of correspondence between energy
densities of $f(T)$ and HDE model. In flat FRW universe, the IR
cutoff $L$ in Eq.(\ref{1}) becomes the future event horizon
$R_{h}=a\int_{t}^{\infty}\frac{dt}{a}$ resulting the HED energy
density. Using Eq.(\ref{a}) in the correspondence, $\alpha$ takes
the form
\begin{equation}\label{30}
\alpha=c^{2}\left(1+\frac{1}{h}\right)^2,
\end{equation}
and $\beta$ is an integration constant. Replacing $f(T)$ and viscous
models in Eq.(\ref{21}), the viscous EoS parameter takes the form
\begin{eqnarray}\nonumber
\omega_{eff}&=&\frac{\kappa^2
\rho_{m0}(1+z)^{3+\frac{3}{h}}+3h\kappa^{2}[\xi_0(1+z)^{2/h}+\xi_{1}h(1+z)^{1/h}+\xi_{2}h(h+1)]}{(\kappa^2
\rho_{m0}(1+z)^{3+\frac{2}{h}}+3\alpha
h^2)(1+z)^{\frac{1}{h}}}\\\label{31}&-&\frac{2\alpha h}{\kappa^2
\rho_{m0}(1+z)^{3+\frac{2}{h}}+3\alpha h^2}-1.
\end{eqnarray}
\begin{figure}
\epsfig{file=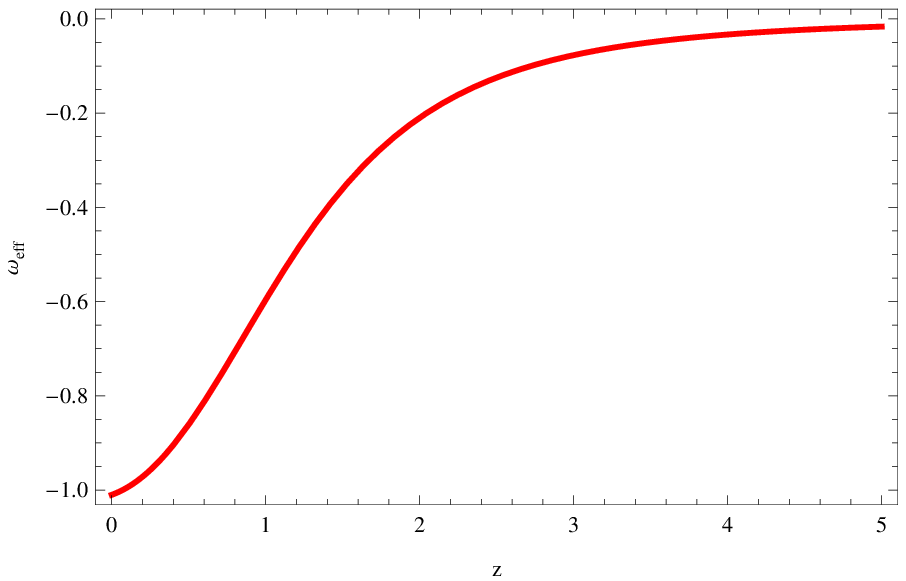, width=0.5\linewidth}\epsfig{file=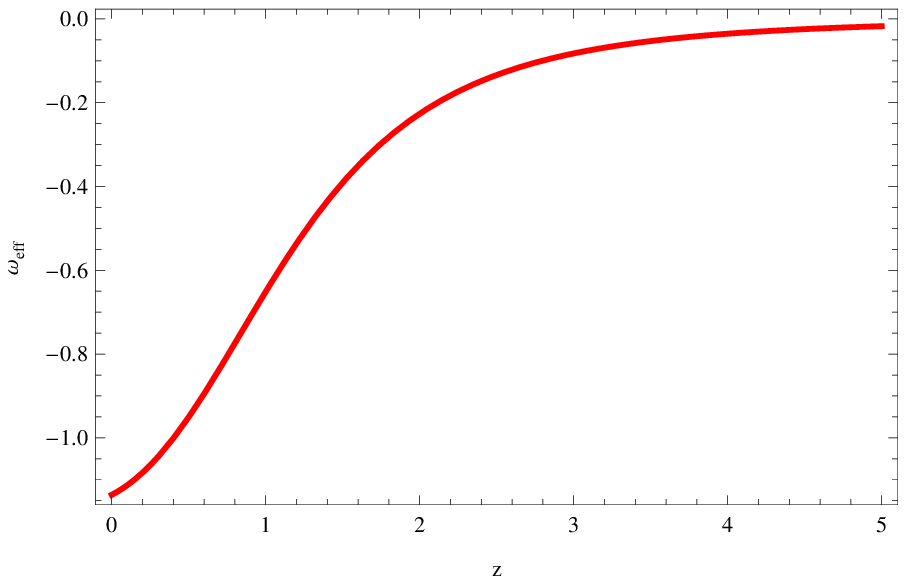,
width=0.5\linewidth}\caption{Plot of time dependent viscous
$\omega_{eff}$ versus $z$ for first model with
$c=0.818,~h=2,~\kappa^2=1=\rho_{m0}$. In the left graph, we take
$\xi_0=0.005,~\xi_1=0.1=\xi_2$ and in the right graph,
$\xi_0=0.005,~\xi_1=0.05=\xi_2$.}
\end{figure}

The graphical behavior of time dependent viscous EoS parameter with
respect to redshift is shown in \textbf{Figure 1}. We draw this
parameter by taking arbitrary values of the coefficients
$(\xi_0,~\xi_1,~\xi_2)$ of viscous model, where $\alpha$ depends
upon the constant $c$ which is 0.818 for flat model \cite{26}. Also,
we fix the redshift range from 0 to 5 to discuss the behavior of the
universe at low redshifts. The left graph in \textbf{Figure 1} shows
the evolution of the universe initially from matter dominated era
for higher values of $z$ and then converges to quintessence era at
$z=1.6$ for $\xi_0=0.005$ and $(\xi_1,~\xi_2)=0.1$. The phantom
divide line is being crossed by the $\omega_{eff}$ as $z$ approaches
to zero. By decreasing $\xi_1$ and $\xi_2$ from $0.1$, the universe
remains in phantom dominated era (shown in the right graph).

For the constant viscous case, we take $\xi_{1}=0=\xi_{2}$ in
Eq.(\ref{24}), thus the constant viscous EoS parameter becomes
\begin{eqnarray}\label{33}
\omega_{eff}&=&\frac{\kappa^2
\rho_{m0}(1+z)^{3+\frac{2}{h}}+3h\kappa^{2}\xi_0(1+z)^{\frac{1}{h}}-2\alpha
h}{\kappa^2 \rho_{m0}(1+z)^{3+\frac{2}{h}}+3\alpha h^2}-1.
\end{eqnarray}
\textbf{Figure 2} represents the same behavior as indicated by time
dependent viscous EoS parameter. However, the phantom crossing for
the constant viscous coefficient occurs at $\xi_0=0.82$, it shows
phantom behavior for $\xi_0<0.82$ (right graph).
\begin{figure}
\epsfig{file=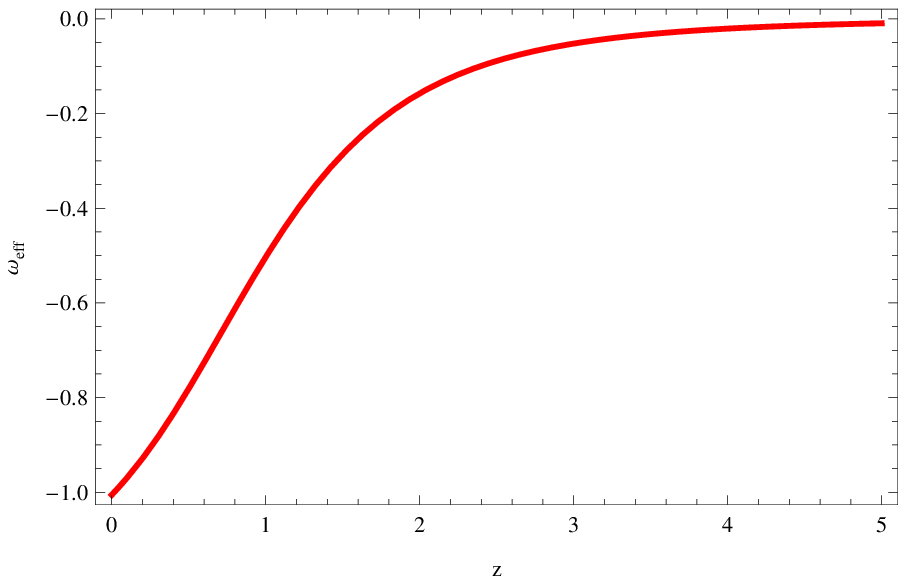, width=0.5\linewidth}\epsfig{file=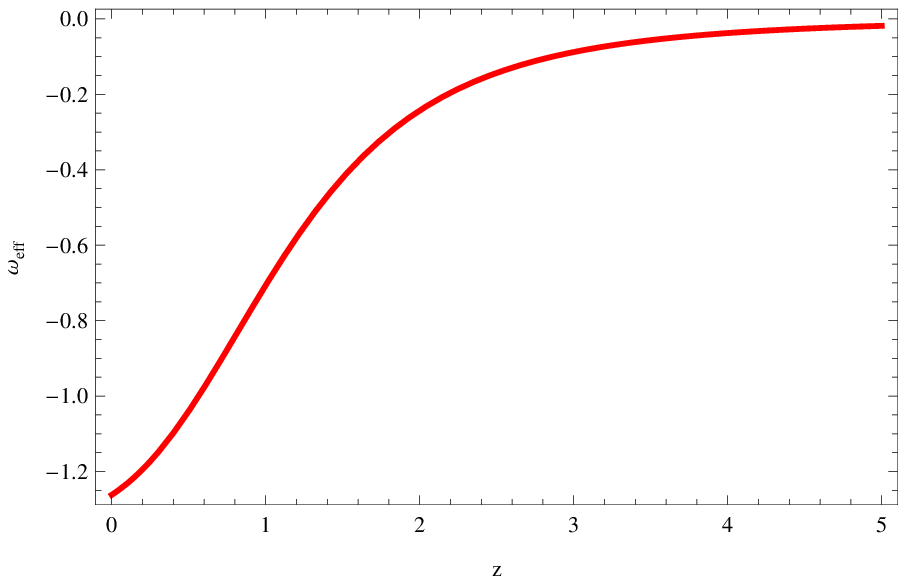,
width=0.5\linewidth}\caption{Plot of constant viscous $\omega_{eff}$
versus $z$ for first model with
$c=0.818,~h=2,~\kappa^2=1=\rho_{m0}$. In the left graph,
$\xi_0=0.82$ and in the right graph, $\xi_0=0.05$.}
\end{figure}

\subsection{The Second Model}

Assuming the exponential $f(T)$ model \cite{a,a+}
\begin{equation}\label{c}
f(T)=Te^{bT},
\end{equation}
where $b$ is an arbitrary constant. Inserting $f(T)$ and viscous
models in Eq.(\ref{21}), the viscous EoS parameter takes the form
\begin{eqnarray}\nonumber
\omega_{eff}&=&-1+\left[2\kappa^{2}\rho_{m0}(1+z)^3+\frac{6h\kappa^{2}}{(1+z)^{\frac{1}{h}}}
(\xi_{0}+\frac{\xi_{1}h}{(1+z)^{\frac{1}{h}}}+\frac{\xi_2h(1+h)}{(1+z)^{\frac{2}{h}}})\right.
\\\nonumber
&+&\left.\frac{4h}{(1+z)^{\frac{2}{h}}}\left(-\frac{6h^2}{(1+z)^{\frac{2}{h}}}
\exp(-\frac{6h^2b}{(1+z)^{\frac{2}{h}}})
(2b^2+5b-\frac{(1+z)^{\frac{2}{h}}}{6h^2})\right.\right.\\\label{d}&-&\left.\left.1\right)\right]
\left[2(1+z)^3+\frac{36h^4}{(1+z)^{\frac{4}{h}}}\exp(-\frac{6h^2b}{(1+z)^{\frac{2}{h}}})\right]^{-1}.
\end{eqnarray}
Figure \textbf{3} represents the graphical behavior of time
dependent viscous $w_{eff}$ versus $z$. In the left graph, the plot
shows the evolution of the universe from matter to DE phase for
higher values of redshift, approximately for $z>2.37$. At $z=2.37$
for particular values $\xi_0=0.005$ and $(\xi_1,\xi_2)=0.2$, the EoS
parameter indicates the quintessence era and approaches to $-1$ as
$z\rightarrow 0$. As we decrease the values of $\xi_1,\xi_2$, the
$\omega_{eff}$ represents the phantom era of the universe.
\begin{figure}
\epsfig{file=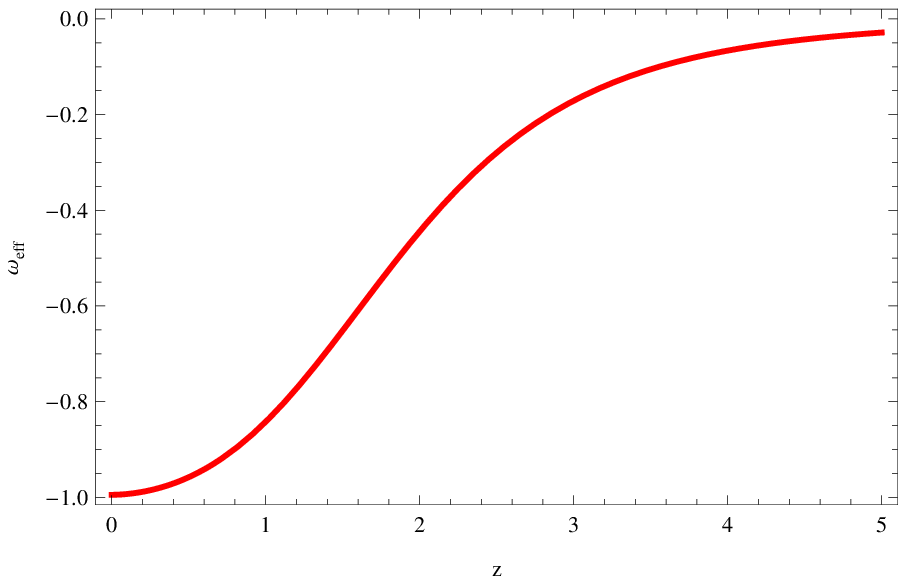, width=0.5\linewidth}\epsfig{file=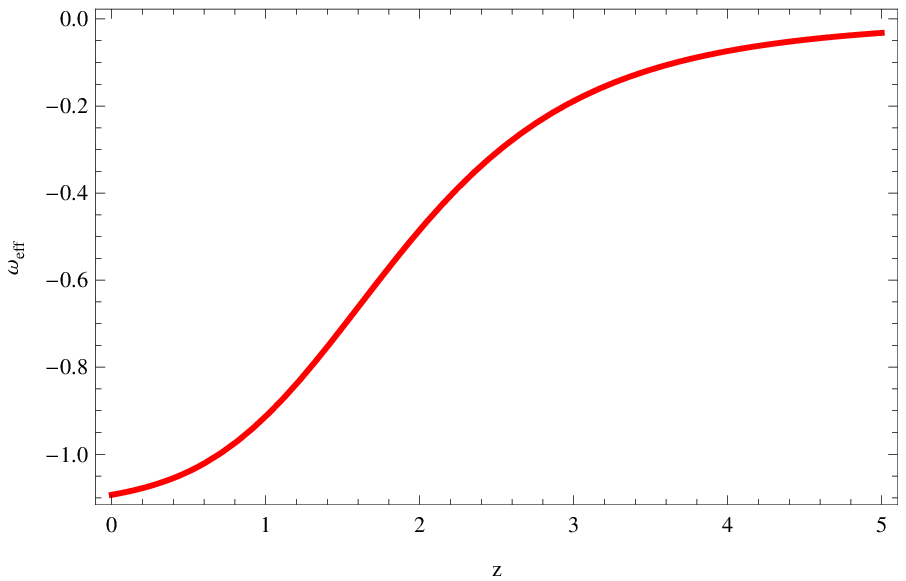,
width=0.5\linewidth}\caption{Plot of time dependent viscous
$\omega_{eff}$ versus $z$ for exponential model with
$b=0.05,~h=2,~\kappa^2=1=\rho_{m0}$. In the left graph, we take
$\xi_0=0.005,~\xi_1=0.2=\xi_2$ and in the right graph,
$\xi_0=0.005,~\xi_1=0.02=\xi_2$.}
\end{figure}

Now for constant viscous EoS parameter, we take $(\xi_1,\xi_2)=0$ in
Eq.(\ref{d}) yields
\begin{eqnarray}\nonumber
\omega_{eff}&=&-1+\left[2\kappa^{2}\rho_{m0}(1+z)^3+\frac{6h\kappa^{2}\xi_{0}}{(1+z)^{\frac{1}{h}}}
+\frac{4h}{(1+z)^{\frac{2}{h}}}\left(-\frac{6h^2}{(1+z)^{\frac{2}{h}}}\right.\right.\\\nonumber&\times&\left.\left.
\exp(-\frac{6h^2b}{(1+z)^{\frac{2}{h}}})
(2b^2+5b-\frac{(1+z)^{\frac{2}{h}}}{6h^2})-1\right)\right]
\left[2(1+z)^3\right.\\\label{e}&+&\left.\frac{36h^4}{(1+z)^{\frac{4}{h}}}
\exp(-\frac{6h^2b}{(1+z)^{\frac{2}{h}}})\right]^{-1}.
\end{eqnarray}
Its plot versus $z$ is in Figure \textbf{4}, showing same behavior
as that of time dependent case. Approximately, the universe meets
the quintessence era at $z<2.1$ and converges towards
$\omega_{eff}=-1$ as $z$ approaches to zero (in left graph). In
right graph, the evolution of EoS parameter represents the phantom
era of the universe for $z\leq0.5$ by decreasing the value of
$\xi_0$, i.e., $\xi_0<1.6$.
\begin{figure}
\epsfig{file=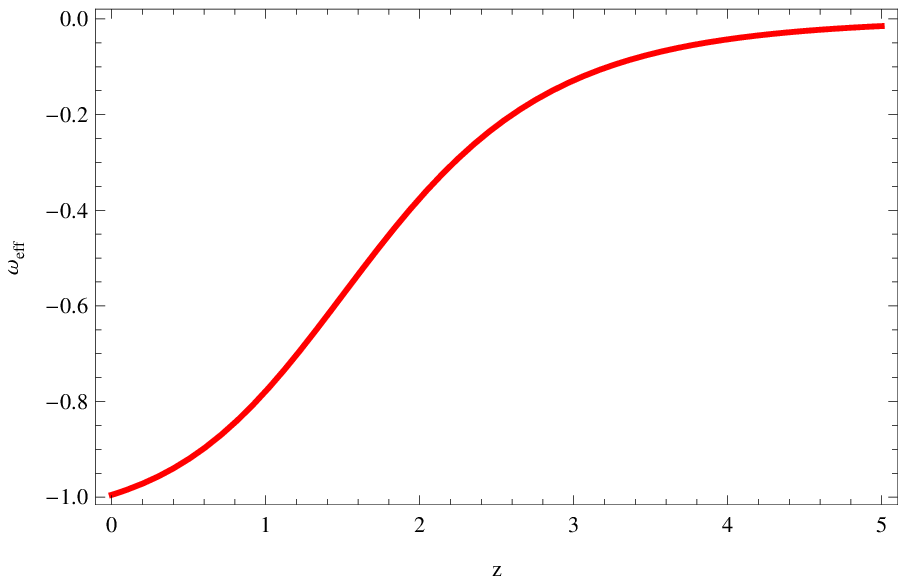, width=0.5\linewidth}\epsfig{file=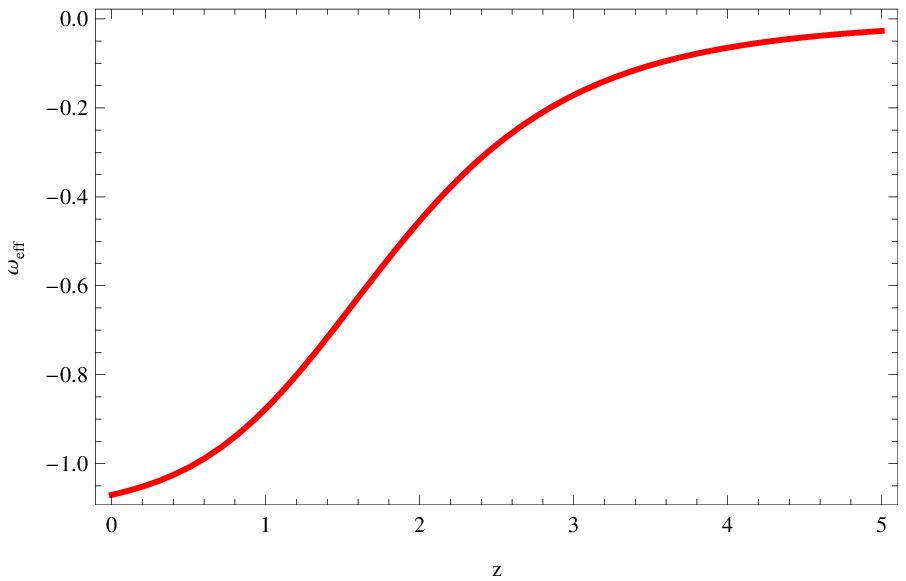,
width=0.5\linewidth}\caption{Plot of constant viscous $\omega_{eff}$
versus $z$ for exponential model with
$b=0.05,~h=2,~\kappa^2=1=\rho_{m0}$. In the left graph, $\xi_0=1.6$
and in the right graph, $\xi_0=0.5$.}
\end{figure}

\subsection{The Third Model}

Finally, we take the model
\begin{equation}\label{38}
f(T)=\epsilon\sqrt{T}+T+\frac{\gamma}{1+2h}T^{1+h},
\end{equation}
which includes linear and nonlinear terms of torsion scalar and
$\epsilon,~\gamma$ are constants. Similar to the first model
(\ref{29}), this model comes through the correspondence of NADE
model with $f(T)$ gravity. The energy density of the NADE model
inherits the conformal time $\eta=\int_{t}^{t_s}\frac{dt}{a}$.
Incorporating the correspondence, here $\epsilon$ is an integration
constant and $\gamma$ is
\begin{equation}\label{40}
\gamma=\frac{6n^2a_{0}^{2}(1+h)^2}{(-6h^2)^{1+h}},
\end{equation}
where $n=2.716$ for flat universe. Replacing Eq.(\ref{38}) in
(\ref{21}), the viscous EoS parameter becomes
\begin{eqnarray}\nonumber
\omega_{eff}&=&\frac{2\kappa^2
\rho_{m0}(1+z)^{3+\frac{3}{h}}+6h\kappa^{2}[\xi_0(1+z)^{2/h}+\xi_{1}h(1+z)^{1/h}+\xi_{2}h(h+1)]}{(2\kappa^2
\rho_{m0}(1+z)^{5+\frac{2}{h}}+\gamma
(-6h^2)^{1+h})(1+z)^{\frac{1-2h}{h}}}
\end{eqnarray}
\begin{eqnarray}\label{41}
&+&\frac{4\gamma
h(1+h)(-6h^2)^{h}(1+z)^{2-h}}{2\kappa^2
\rho_{m0}(1+z)^{5+\frac{2}{h}}+\gamma (-6h^2)^{1+h}}-1.
\end{eqnarray}
\begin{figure}
\epsfig{file=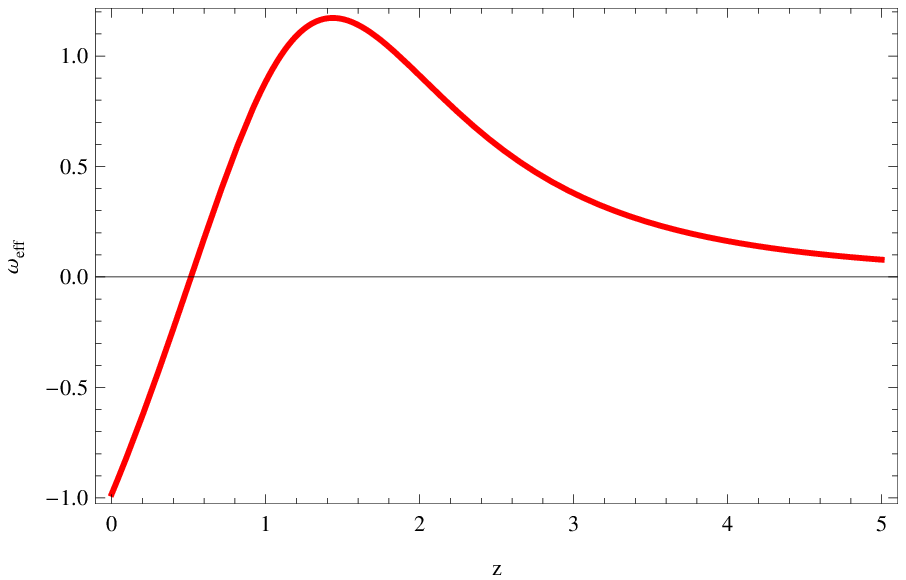, width=0.5\linewidth}\epsfig{file=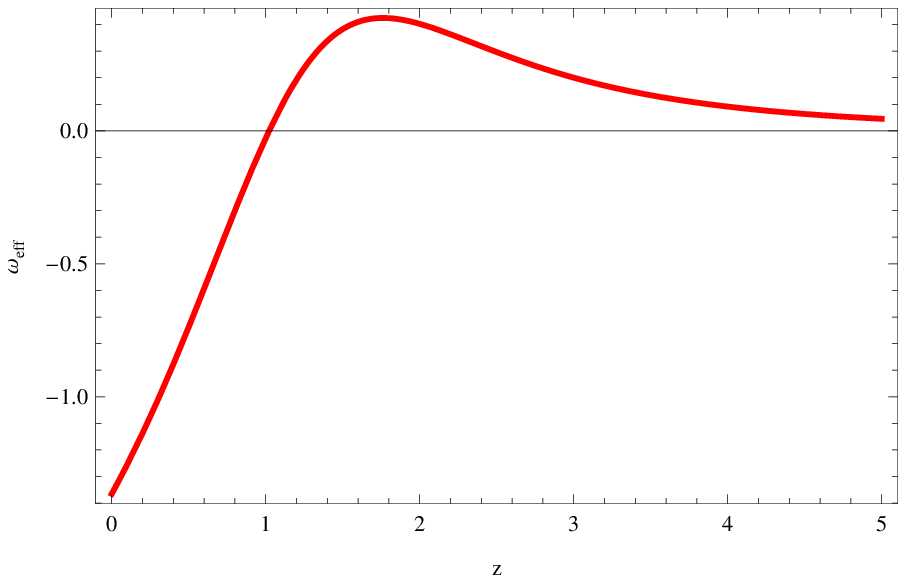,
width=0.5\linewidth}\caption{Plot of time dependent viscous
$\omega_{eff}$ versus $z$ for third model with
$n=2.716,~h=2,~a_0=1,~\kappa^2=1=\rho_{m0}$. In the left graph,
$\xi_0=0.05,~\xi_1=4.2=\xi_2$ and in the right graph
$\xi_0=0.05,~\xi_1=2.6=\xi_2$.}
\end{figure}

The graphical behavior of time dependent viscous $\omega_{eff}$ is
given in \textbf{Figure 5}. Initially, it shows the deceleration
phase $(\omega_{eff}>-\frac{1}{3})$ of the universe for higher
values of $z$. As we decrease the value of redshift up to $0.4$, it
meets the quintessence region for the particular values $\xi_0=0.05$
and $(\xi_1,~\xi_2)=4.2$, and crossing of the phantom divide line
takes place for $z$ tends to zero. The right graph indicates that
the universe remains in this era for $(\xi_1,~\xi_2)<4.2$. The
constant viscous model for this case is
\begin{eqnarray}\nonumber
\omega_{eff}&=&\frac{2\kappa^2
\rho_{m0}(1+z)^{5+\frac{2}{h}}+6h\kappa^{2}\xi_0(1+z)^{2+\frac{1}{h}}}{2\kappa^2
\rho_{m0}(1+z)^{5+\frac{2}{h}}+\gamma (-6h^2)^{1+h}}-1\\\label{43}
&+&\frac{4\gamma h(1+h)(-6h^2)^{h}(1+z)^{2-h}}{2\kappa^2
\rho_{m0}(1+z)^{5+\frac{2}{h}}+\gamma (-6h^2)^{1+h}}.
\end{eqnarray}
\textbf{Figure 6} shows its plot versus redshift. It provides the
crossing of phantom divide line for a high value $\xi_0=32$, whereas
$\xi_0\leq32$ corresponds to the phantom region for decreasing $z$
of the accelerating expansion of the universe.
\begin{figure}
\epsfig{file=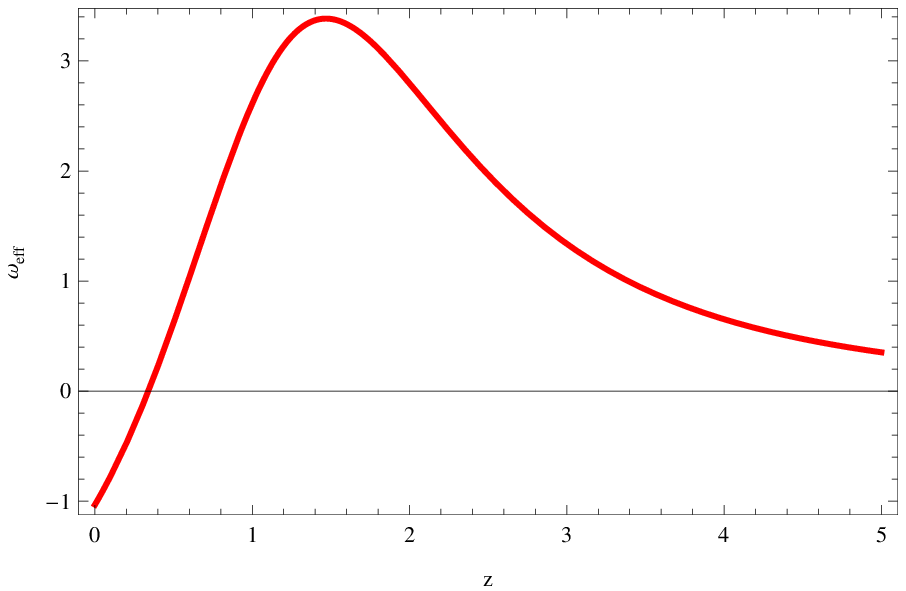, width=0.5\linewidth}\epsfig{file=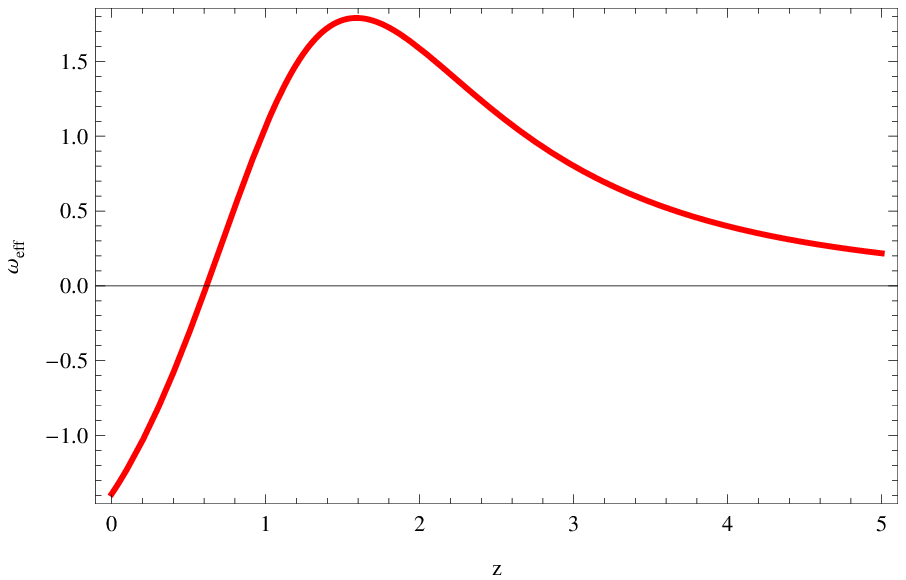,
width=0.5\linewidth}\caption{Plot of constant viscous $\omega_{eff}$
versus $z$ for third model with
$n=2.716,~h=2,~a_0=1,~\kappa^2=1=\rho_{m0}$. In the left graph,
$\xi_0=32$ and in the right graph $\xi_0=20$.}
\end{figure}

\section{Outlook}

Viscous models have been discussed in cosmological evolution of the
universe as compared to the ideal perfect fluid. The term of shear
viscosity vanished when a completely isotropic unverse is assumed
and only the bulk viscosity contributes for the accelerating
universe to get negative pressure. In this paper, we have considered
viscosity by taking dust matter in the framework of $f(T)$ gravity.
We have taken three different viable DE models and a time dependent
viscous model to construct the viscous EoS parameter for these
models. The graphical representation is also developed by
considering arbitrary values of the coefficients in viscous model
for a specific expression of Hubble parameter. The results and the
comparison with non-viscous case are given as follows.

All the three models in viscous fluid indicates the behavior of the
universe from matter dominated phase to quintessence era and then
converges to phantom era of the DE dominated phase for decreasing
$z$. It shows the phantom universe by taking the particular values
of viscous coefficients. The constant viscous cases also exhibit
phantom behavior. The non-viscous case $\xi=0$ shows a universe
which always stays in phantom for $h>0$ or quintessence for $h>1$
regions \cite{26}. However, the third model has resulted the phantom
phase of the universe for the higher values of viscous coefficients
as compared to the first and second $f(T)$ models. In each case, the
time dependent case shows the phantom crossing by taking small
values of viscous coefficients while constant viscous case needs
higher values for crossing.

The combination of torsion and viscosity influences the accelerating
expansion of the universe in such a way that it strictly depends
upon the viscous coefficients of the model. We have to fix the
ranges for these coefficients in order to get our desired results.
We conclude that the viscosity model leads to different behavior of
the accelerating universe in DE era under the effects of viscous
fluid. On the other hand, viscosity may result the crossing of the
phantom divide line and phantom dominated universe \cite{6,7,8} as
shown in \textbf{Figures 1} and \textbf{6}. In the non-viscous case
\cite{26}, the universe remains in the phantom and quintessence eras
for the relevant scale factors. Beyond the ideal situation, we
remark that the DE era of the universe in a real fluid may be
observed and hence accelerating expansion of the universe is
achieved.

\end{document}